\documentclass{svmult}

\usepackage{makeidx}     
\usepackage{graphicx}    
\usepackage{multicol}    

\makeindex             

\begin{document}

\title*{Simulation of Vortex--Dominated Flows Using the FLASH Code}
\author{Vikram Dwarkadas,\inst{1} Tomek Plewa,\inst{1} Greg Weirs,\inst{1}
 Chris Tomkins,\inst{2} and Mark Marr-Lyon\inst{2}}
\authorrunning{Dwarkadas, Plewa, Weirs, Tomkins, Marr-Lyon}
\institute{ASCI FLASH Center, University of Chicago
\texttt{vikram@flash.uchicago.edu, tomek@flash.uchicago.edu, weirs@flash.uchicago.edu}
\and Los Alamos National Laboratory \texttt{ctomkins@lanl.gov, mmarr@lanl.gov}}
%
%
\maketitle

\section{Abstract}

We compare the results of two--dimensional simulations to experimental
data obtained at Los Alamos National Laboratory in order to validate
the FLASH code.  FLASH is a multi--physics, block--structured adaptive
mesh refinement code for studying compressible, reactive flows in
various astrophysical environments.  The experiment involves the
lateral interaction between a planar Ma=1.2 shock wave with a cylinder
of gaseous sulfur hexafluoride (SF$_6$) in air.  The development of
primary and secondary flow instabilities after the passage of the
shock, as observed in the experiments and numerical simulations, are
reviewed and compared.  The sensitivity of numerical results to
several simulation parameters are examined.  Computed and
experimentally measured velocity fields are compared.  Motivation for
experimental data in planes parallel to the cylinder axis is provided
by a speculative three--dimensional simulation.


\section{Introduction}

The impulsive acceleration of a material interface can lead to complex
fluid motions due to the Richtmyer--Meshkov (RM) instability.  Here,
the misalignment of pressure and density gradients deposits vorticity
along the interface, which drives the flow and distorts the interface.
At later times the flow may be receptive to secondary instabilities,
most prominently the Kelvin--Helmholtz instability, which further
increase the flow complexity and may trigger transition to turbulence.

Experimental investigations of impulsively accelerated interfaces have
focused mainly on interfaces with single--mode perturbations, and on
the case we consider here, shock--accelerated cylindrical gas
columns~\cite{jacobs93,tompre03,zoldi2002}.  The experiments are
relatively inexpensive and turnaround times are short.  The challenges
are diagnostics and repeatability: after the shock passes, the
flowfield evolution is driven by flow instabilities and vortex
dynamics, which are sensitive to the initial conditions and noise in
the system.  Specifically, for the present case of a single
shock--accelerated gas cylinder, the flow is dominated by a
counter--rotating vortex pair.

Verification and validation are critical in the development of any
simulation code, without which one can have little confidence that the
code's results are meaningful.  The sensitivity and the complex
evolution of the vortex pair are desirable properties for our primary
purpose, which is to use the experiments to validate our simulation
code.  A well--designed, well--characterized, and accurately diagnosed
experiment is essential for validation.

FLASH is a multi--species, multi--dimensional, parallel,
adaptive--mesh--refinement, fluid dynamics code for applications in
astrophysics~\cite{fryols00}.  Calder et al. discuss initial
validation tests of the FLASH code~\cite{calfry02}.  Herein we
continue our validation effort by comparing FLASH simulations to an
experiment performed at the Los Alamos National
Laboratories~\cite{tompre03,zoldi2002}.

\section{Two--Dimensional Simulations}

\subsection{Experimental Facility and Initial Conditions}

The experimental apparatus is a shock--tube with a 7.5~cm square
cross--section.  Gaseous SF$_6$ flows from an 8~mm diameter nozzle in
the top wall of the shock--tube, forming a cylinder of dense gas in
the otherwise air--filled test section.  A Ma=1.2 shockwave travels
through the shock--tube and passes through the cylinder. Our interest
is in the resulting evolution of the SF$_6$. The experiment is
nominally two--dimensional, and the experimental data are taken in a
plane normal to the cylinder axis.

The initial SF$_6$ distribution (before the shock impact) is
visualized by Rayleigh--Scattering from the SF$_6$
molecules~\cite{tompre03}.  The image plane is 2~cm below the top wall
of the test section.  As the SF$_6$ flows downward, air diffuses into
the SF$_6$ column, reducing the peak concentration of the heavy gas.
One limitation of the visualization technique is that the pixel
intensity in the images gives only the mole fraction of SF$_6$
relative to the peak mole fraction.  The scaling between pixel
intensity and mole fraction is linear, with the proportionality
constant specified by the maximum initial mole fraction of SF$_6$,
denoted X$_{SF6}$.

After shock passage, two--dimensional velocity vectors in a plane are
obtained using particle image velocimetry (PIV)~\cite{prevor00}.  The
technique yields high resolution (spacing between vectors is about 187
microns) and high accuracy (measurement error is 1.5\% of the
structure convection velocity).  Raw images are interrogated using a
two-frame cross-correlation technique~\cite{chrsol00}, which produces
approximately 3600 vectors per realization.  For PIV both the air and
the SF$_6$ must be seeded with water/glycol droplets, nominally 0.5
$\mu$m in diameter, the displacement of which is used to obtain
velocity data---hence, simultaneous velocity and composition images
cannot be obtained.

\begin{figure}
\centering
\hfil
\includegraphics[height=4.8cm, angle=-90]{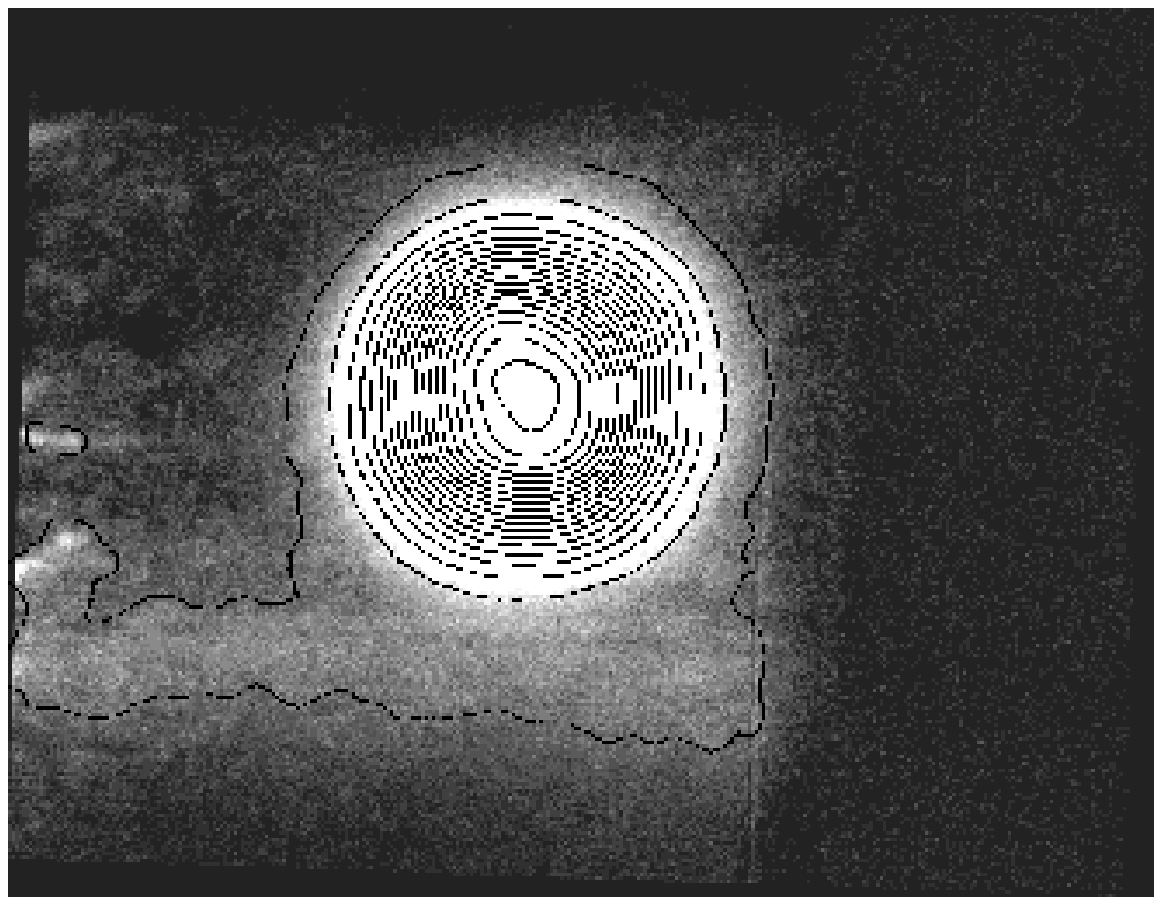}
\hfil
\includegraphics[height=4.8cm, angle=-90]{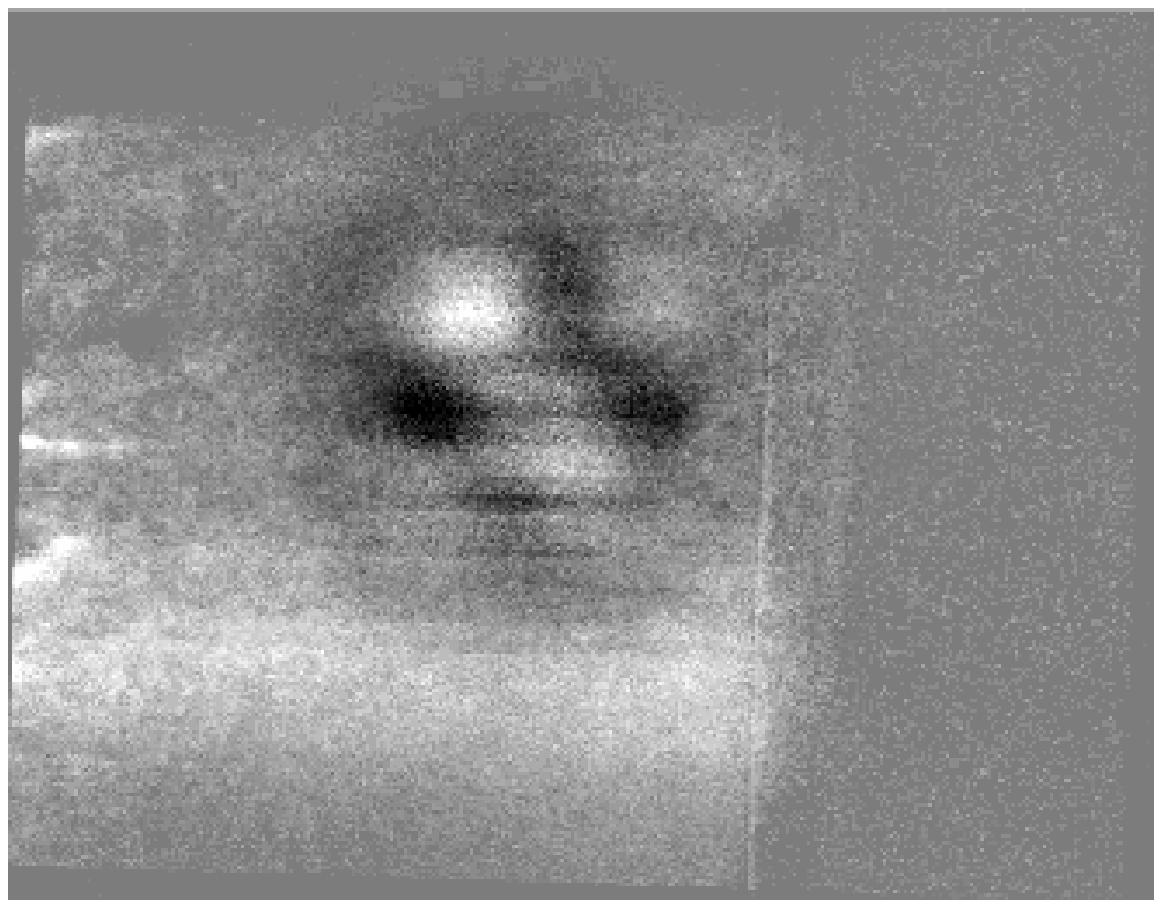}
\hfil
\caption{(a) The initial conditions for the single--cylinder experiment
with an 8mm nozzle. (b) Residuals between the experimental image and
the composite Gaussian fit.  The scale varies from -10 (black) to +10
(white) in intensity units.  The maximum signal in the image has an
intensity of about 165.  Note a semi--regular non--radial m=4,l=4
component with maximum signal reaching about 10\% of local intensity.}
\label{f:cyl_image}
\end{figure}

The $608 \times 468$~pixel image of the initial SF$_6$ distribution is
shown in Fig.~\ref{f:cyl_image}.  The pixel size is 38~microns when
projected into the measurement plane.  The pixel intensity is plotted,
with 20 contours equally spaced between values of 5 and 165.  The
deformation of the contours indicates that the distribution of SF$_6$,
as revealed by the diagnostics, is only approximately radially
symmetric.  Also, the signal is completely dominated by noise at the
level of about 5--10\% (the two lowest density contours in
Fig.~\ref{f:cyl_image}).  Since the asymmetries are likely to vary
from one experimental shot to another and the flowfield evolution is
highly sensitive to noise, smooth initial conditions for our
simulations are obtained by fitting a radially--symmetric function to
the experimental data.

To obtain the smooth initial conditions we used the MINGA minimization
package~\cite{plewa88}.  The fit extended out to a radius of 150
pixels from the center.  After examining many trial functions, we
selected the form:
\[
C(r)  = u_1 e^{(-r^2/u_2^2)}
      + u_3 e^{(-r^2/u_4^2)}
      + u_5 e^{(-(r-u_6)^2/u_7^2)}
      + u_8 e^{(-(r-u_9)^2/u_{10}^2)}
\]
where $u_1 = 144.0725$, $u_2 = 69.45422$, $u_3 = 9.221577$, $u_4 =
20.10299$, $u_5 = 32.47960$, $u_6 = 42.59238$, $u_7 = 32.10067$, $u_8
= -1.559247$, $u_9 = 98.27106$, and $u_{10} = 15.51837$.  (The length
units are pixels, and the maximum intensity will be rescaled to
X$_{SF6}$.)  Residuals are shown in Fig.~\ref{f:cyl_image}.  The
experimental data appears to contain a significant non--radial signal
which can be characterized by an m=4, l=4 perturbation with an
amplitude of about 10\%. Our fit does not account for this additional
component.

\subsection{Overview of the Simulations}

As the shock travels, it accelerates the medium through which it
propagates.  As the shock traverses the cylinder, vorticity is
deposited baroclinically along the interface, i.e., due to the
misalignment of the pressure gradient (normal to the shock) and the
density gradient (normal to the interface.)  The vorticity deposition
is not uniform: it is maximum when the gradients are perpendicular,
and since the shock is slowed in the SF$_6$, the maximum is shifted to
the downstream portion of the cylinder edge.  Once the shock has
passed through the SF$_6$, vorticity generation due to the primary
instability is complete.  The existing vorticity drives the flow: a
counter--rotating vortex pair forms, then rolls up.  The vortex
Reynolds number of the flow, as measured experimentally, is Re$=\Gamma
/ \nu \approx 5 \times 10^4$.  The development and evolution of the
vortex pair and subsequent instabilities at the interface proceed in a
weakly compressible regime.  More precise descriptions can be found in
the references~\cite{jacobs93,quikar96,zoldi2002}.

\begin{figure}
\centering
\includegraphics[scale=0.65,clip=true]{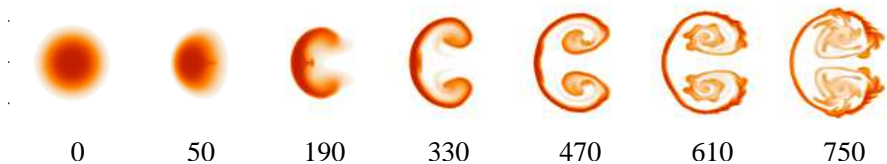}
\caption{Evolution of the SF$_6$, with time elapsed after shock impact
listed in $\mu$s.  The mass fraction of SF$_6$ is shown, with
X$_{SF6}$=0.6 and CFL=0.8.}
\label{f:series}       
\end{figure}

Figure~\ref{f:series} shows a sequence of images from our baseline
simulation.  The minimum grid resolution is 78~microns, the initial
peak mole fraction of SF$_6$ is~0.6, and the Courant (CFL) number
is~0.8.  (The CFL number is a nondimensional measure of the timestep
size.)  For all simulations, the streamwise and spanwise extent of the
domain were~64 cm and 8 cm, respectively.  Overall the flow features
in the simulation results are similar to those in the experimentally
obtained images.  Next we will describe the effects of several
simulation parameters on the computed results.  The amount and
location of small--scale structure, relative to the experimental data,
is used as a qualitative metric.

\subsection{Effects of Simulation Parameters}
\subsubsection{Effect of Maximum Initial Mole Fraction}

\begin{figure}
\centering
\includegraphics[height=3.5cm,clip=true]{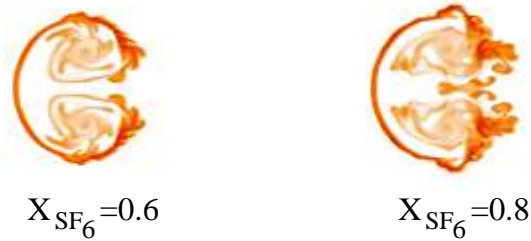}
\caption{SF$_6$ mass fraction at 750$\mu$s after shock impact.
Initial maximum SF$_6$ mole fraction X$_{SF6}$=0.6 on the left, and on
the right X$_{SF6}$=0.8.}
\label{f:vol_frac}       
\end{figure}

Because the experimental images provide only information about the
relative mole fractions, the maximum mole fraction of SF$_6$ at the
start of the simulation is a free parameter in our initial conditions.
We have focused on two values, X$_{SF6}$=0.8, motivated by
experimental estimates, and X$_{SF6}$=0.6, better supported by
comparison of simulation and experimental results.  Simulation results
are presented in Fig.~\ref{f:vol_frac}.  For X$_{SF6}$=0.8, the
numerical solution shows excessive small--scale structure compared to
the experimental data.

\subsubsection{Effect of Grid Resolution}

\begin{figure}
\centering
\includegraphics[height=3.5cm, clip=true]{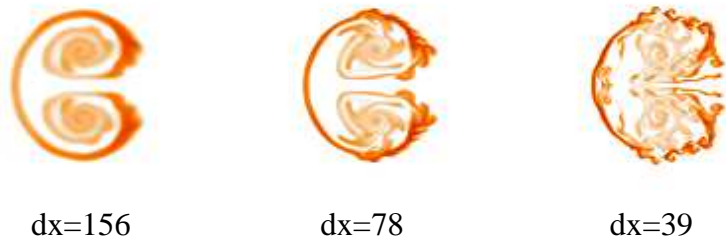}
\caption{SF$_6$ mass fraction at 750~$\mu$s with increasing
grid resolution, labeled in microns.  At the
highest level of resolution, the vortex cores appear diffuse.}
\label{f:resol}       
\end{figure}

Using FLASH's adaptive mesh refinement capability, we have run
simulations at three grid resolutions.  Results at 750~$\mu$s are
presented in Fig.~\ref{f:resol}, with minimum grid spacings of
156~microns, 78~microns and 39~microns. We observe that the amount of
small--scale structure increases on finer grids.  This can be
understood since the numerical dissipation in FLASH's shock--capturing
scheme (PPM) is resolution dependent, and no physical viscosity model
was used for these simulations.  (Estimates of the length scale of
molecular diffusion at the flow conditions of the experiment are below
10~microns.)  At a grid resolution of 39~microns, the primary vortex
cores are not easily identified among the diffuse, turbulent
structures.  At the lower resolutions, the two primary vortex cores
are unambiguous.

\subsubsection{Flow--Mesh Interaction}

It is known that unavoidable interpolation errors near discontinuous
jumps in grid resolution can act as sources of spurious small--scale
structure~\cite{quirk1991}.  To test this possibility we have run
simulations in which a predetermined area around the vortices is
uniformly refined to the highest resolution.  Compared to fully
adaptive refinement (the default) this approach significantly reduces
the amount of perturbations introduced by the grid adaption process.

\begin{figure}
\centering
\includegraphics[height=3.5cm, clip=true]{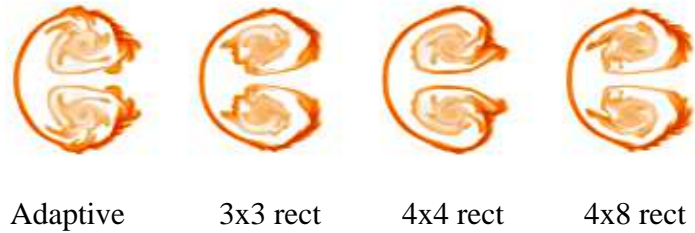}
\caption{Solutions on different grids, 750~$\mu$s after shock impact
at CFL=0.8.  Left to right: fully adaptive grid; $3 \times 3$~cm
refined rectangle; $4 \times 4$~cm refined rectangle; $4 \times 8$~cm
refined rectangle.  In the rightmost image, the refined rectangle
covers the entire spanwise extent of the test section.}
\label{f:rect_ref}       
\end{figure}

In Fig.~\ref{f:rect_ref} we compare the results from a fully adaptive
grid and grids with maximally refined rectangles of $3 \times 3$~cm,
$4 \times 4$~cm, and $4 \times 8$~cm.  The vortex structure is always
less than 2~cm across.  The results are shown at 750~$\mu$s after
shock impact.  For all grids, the minimum grid spacing is 78~microns
and the CFL number is 0.8.  For the different grids the large scale
morphology, such as size of the cylinder cross--section and the basic
vortex structure, remains the same.  However, the shape of the
cross--section visibly differs depending on the grid, as does the
amount and location of small--scale structures.  In particular,
differences are noticeable in the small--scale instabilities present
on the vortex rolls.  Since all other simulation aspects are the same,
the differences must originate with perturbations at jumps in
refinement.

\subsubsection{Effect of Courant Number}

The timestep in an explicit hydrodynamic code is limited by the CFL
number.  In general we use a value of CFL=0.8, but we have also
performed simulations with the time step size limited by CFL=0.2.
Reducing the timestep generally reduces the temporal truncation error;
however, it might have an adverse effect on the spatial error.
Additionally the mesh can adapt more often per unit of simulation
time, and if errors are committed every time the mesh adapts, this
could lead to a less accurate solution.

\begin{figure}
\centering
\includegraphics[height=3.5cm, clip=true]{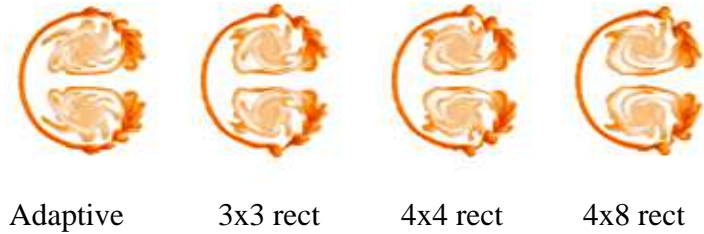}
\caption{Solutions on different grids for CFL=0.2; for other details
see Fig.~\ref{f:rect_ref}.}
\label{f:cfl}       
\end{figure}

We repeated the simulations described above, but at CFL=0.2.  The
results are shown in Fig.~\ref{f:cfl}.  The simulations at CFL=0.2 are
much less sensitive to grid adaption: there is much less variation
between solutions on adaptive and locally uniform grids at CFL=0.2
than at CFL=0.8.  One explanation for these results is that the errors
at the fine--coarse boundaries are larger and lead to stronger
perturbations at higher CFL numbers.  An alternative explanation is
that at higher Courant numbers, PPM does not adequately compute
solutions at these conditions.  Our simulations indicate that for
FLASH, a lower CFL number is preferred in this regime.

Our fully adaptive grid simulations provides a speed--up factor of
about~10, compared to the grids with the $4 \times 8$~cm maximally
refined rectangle. Such large savings demonstrate the advantages of
adaptive mesh refinement.  At the same time, caution is warranted: our
results also demonstrate that AMR generates perturbations which,
depending on other simulation parameters and the flow regime, can give
rise to spurious small--scale structures.

\subsection{Metrics for Comparison to Experimental Data}
\subsubsection{Evolution of Cylinder Size}

\begin{figure}
\centering
\includegraphics[scale=0.3]{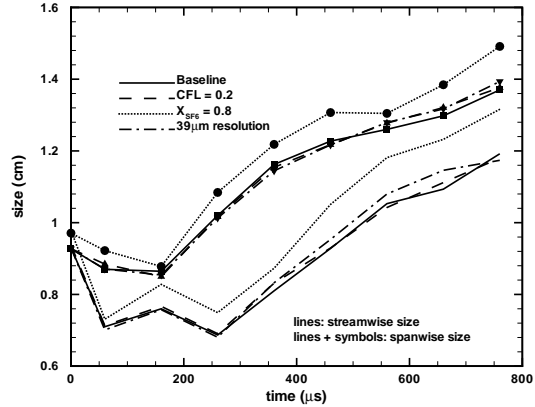}
\caption{Integral scale measures plotted at nine simulation times.
For the baseline case, X$_{SF6}$=0.6, CFL=0.8, and 78 micron
resolution.  For other cases a single parameter varies from the
baseline case.}
\label{f:integrals}       
\end{figure}

In Fig.~\ref{f:integrals} we present integral scale measures -- the
streamwise and spanwise extent of the SF$_6$ -- over time.  The
contour of SF$_6$ mass fraction equal to 0.1 was used to define the
edge of the SF$_6$.  We plot results for our baseline simulation and
simulations where a single parameter is varied relative to the
baseline.  For a given maximum initial SF$_6$ mole fraction, the
cylinder height and width are essentially insensitive to the
parameters varied.  These integral measures provide a basis for
comparison of our simulation results to those of others
(e.g.~\cite{zoldi2002,ridkam02}), as well as to the experimental data.
\subsubsection{Velocity Comparisons}

We compare the magnitude of the velocity fluctuations from the
experimental data to that from the simulation results. The velocity
fluctuation is defined as the velocity in the frame of reference of
the vortices.  To find the velocity fluctuation we subtract the
convective velocity, which we define as the streamwise velocity
component in the lab frame at the point of the maximum vorticity.

\begin{figure}
\centering
\includegraphics[height=5cm]{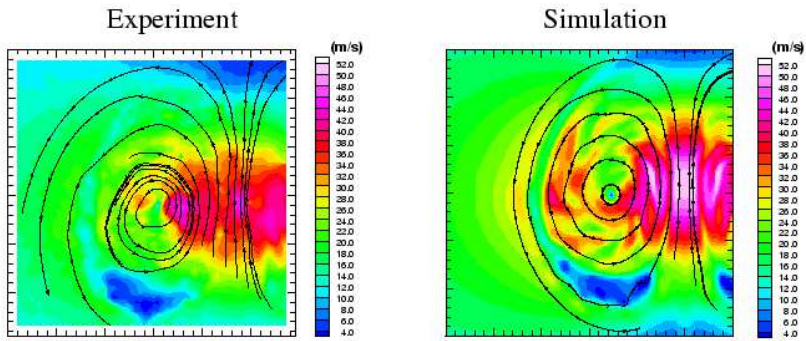}
\caption{ Velocity fluctuation magnitude and streamlines around one primary
vortex 750~$\mu$s after shock impact. The shock has passed through the
image from the top to bottom. Left: experimental data. Right:
simulation data.}
\label{f:velocity}       
\end{figure}

The experimental data are the velocity components and the
corresponding vorticity on $60 \times 60$~points uniformly covering
a $12 \times 12$~mm region around one of the primary vortices.  We
find the convective velocity to be 101.25~m/s.  The magnitude of the
velocity fluctuations are plotted as flooded contours in the left plot
of Fig.~\ref{f:velocity}, together with streamlines of the velocity
fluctuations.  The plot is oriented such that the shock has passed
through the image from top to bottom, and the centerline (between the
primary vortices) is near the right edge of the plot.

The right plot of Fig.~\ref{f:velocity} shows the same quantities but
from the simulation data (78~micron resolution, X$_{SF6}$=0.6,
CFL=0.2), for which the convective velocity is 92.90~m/s.  The
velocity fluctuation field in the simulations compares reasonably well
to the experimental data. The maximum and minimum fluctuations occur
in approximately the same places.  The velocities in the simulation
are up to 10\% higher than in the experiment.  The simulation data
shows more structure, and there is some (visual, at least) correlation
between the structure in the velocity fluctuations and the SF$_6$
distribution.

\section{Three--Dimensional Simulation}

While the experiment is nominally two--dimensional, several factors
might contribute to non--negligible three--dimensional effects.
Probably the most significant experimental deviations from
two--dimensionality are due to the diffusion of the SF$_6$ as it flows
from the top of the test section to the bottom.  As a result, the
mixture composition and density vary with height.  (This is why we
assume the maximum initial mole fraction of SF$_6$ is less than 1.0 in
the two--dimensional simulations.) Even if the cylinder were invariant
with height, the flowfield after the shock passes the cylinder is
dominated by vorticity dynamics, and small scale structures arising
from flow instabilities eventually develop at the vortex edges.  The
effects of such inherently three--dimensional phenomena cannot be
captured in two--dimensional simulations.

We executed a speculative three--dimensional simulation which, though it cannot
definitively establish that three--dimensional effects are essential
for the shock--cylinder interaction, compels further experimental
and computational investigation.  Our z-direction extension of the
initial conditions is purely ad hoc, because we have no corresponding
experimental data.  Of course, this simulation cannot be used as a
validation test for the FLASH code, but we hope it will open a new
line of investigation and discussion.

For this simulation the z-direction is parallel to the cylinder axis,
with z=0.0~cm at the bottom wall and z=8.0cm at the top wall.  To
initialize the flowfield, we begin with the ``raw'' image in
Fig.~\ref{f:cyl_image}.  Only after this simulation was completed did
we learn that the raw images have spurious high frequency noise, and
smooth approximations to the raw data are believed to better represent
the true SF$_6$ field~\cite{tompre03}.  The streamwise (x) and
spanwise (y) dimensions of the image are rescaled linearly with z, so
the SF$_6$ covers a smaller area at the top wall and a larger area at
the bottom.  Consequently the maximum mole fraction of SF$_6$ in each
plane varies as z$^2$, and is~0.64 at the top wall and~0.47 at the
z=6.0~cm plane, at which the ``raw'' image was obtained.  Otherwise
the initialization is the same as for the two--dimensional
simulations.  The simulation was run at CFL=0.8 using fully adaptive
mesh refinement, and the minimum grid spacing was 156~microns in all
three spatial dimensions.

Near the top wall in our simulation, the vortices have rolled up more
than at the bottom, and more small scale structure has developed.
This observation appears to hold throughout the course of the
simulation, and can be understood as follows.  The rescaling of the
initial SF$_6$ distribution results in larger composition gradients,
and correspondingly larger density gradients, near the top and smaller
gradients near the bottom.  The primary source of vorticity generation
in this problem is through baroclinic torque, so more vorticity is
deposited near the upper wall, where the density gradients are
largest.

\begin{figure}
\centering
\includegraphics[scale=0.3]{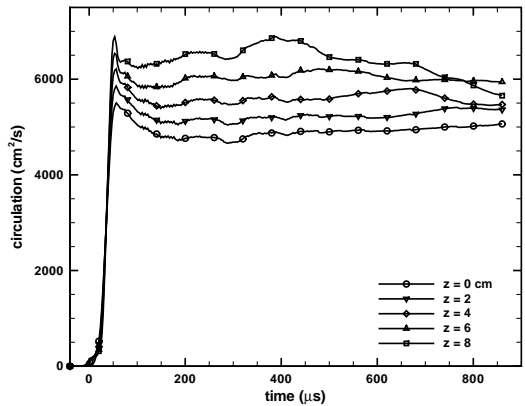}
\caption{Z--component of circulation in the lower-y half of the
xy-plane at five vertical locations in the tunnel, as a function of
time.}
\label{f:circ}       
\end{figure}

The circulation provides a quantitative measure of the vortex
development.  The z-component of circulation was calculated for the
lower-y half of each xy-plane of the simulation, bounded by the
centerline, y=0.0~cm, and the inflow and outflow
boundaries. Figure~\ref{f:circ} shows the circulation at z=0.0, 2.0,
4.0, 6.0 and 8.0~cm as functions of time.  The profiles are
essentially the same up to 300~$\mu$s after the shock impact, with the
magnitude in each plane increasing from the bottom wall to the top
wall; this suggests that there are no significant three--dimensional
effects through this time other than the diffusion of the SF$_6$ as it
flows vertically.  However, after 300~$\mu$s, differences between the
profiles for each height appear, beginning with the profile near the
top wall, and eventually spreading to lower heights.

We find that by the end of the simulation, the z-component of velocity
has reached a maximum of 17~m/s, which is more than half of the
maximum of the spanwise component.  Positive z-velocity is maximum in
the vortex cores, away from the upper and lower walls.  Apparently,
the stronger vortices near the upper wall have correspondingly lower
core pressures compared to the weaker vortices near the lower wall;
this pressure difference accelerates gas toward the upper wall through
the vortex cores.  The acceleration is compounded because the cores of
the vortices are filled with air, and the heavier SF$_6$ is wrapped
around the outside.  The lighter air in the core of each vortex is
inside a tube of SF$_6$, and is preferentially accelerated toward the
upper wall.  The different circulation profiles and large z-component
of velocity suggest that three--dimensional effects are not negligible
for the initial conditions we assumed.  Only with experimental data
can we test those assumptions.

\section{Concluding Remarks}

To date we have made a large number of two--dimensional simulations to
validate the FLASH code for problems dominated by vortex dynamics.
While this is work in progress, we can make the following remarks:
\begin{itemize}
\item 
The overall morphology of the flowfield is captured by the
simulations, but differences exist in the location and extent of
small--scale structure.
\item
Simulation velocity magnitudes lie within 10\% of experimental values.
\item
For vortex--dominated subsonic flows, FLASH users should be cautious
regarding the choice of CFL number, mesh refinement criteria, and if
PPM is used, contact steepening.
\end{itemize}
Our three--dimensional simulation, despite issues with the initial
conditions, suggests that three--dimensional effects might be
important for this experiment, and measurements should be made
parallel to the cylinder axis to examine the issue.

\vspace*{-0.2in}
\bibliography{vikram}
\bibliographystyle{unsrt}

\printindex
\end{document}